\documentclass[aps,prx,twocolumn,showpacs,psfig,superscriptaddress,longbibliography]{revtex4-2}
\usepackage[absolute,overlay]{textpos}
\usepackage{graphicx}
\usepackage{float}
\usepackage{latexsym,amsmath,amssymb,bm,euscript}
\usepackage{color}
\usepackage{subfigure}
\usepackage{epstopdf}
\usepackage[colorlinks=true,linkcolor=blue,citecolor=blue]{hyperref}
\usepackage{soul}
\usepackage[normalem]{ulem}
\usepackage{mathrsfs}
\usepackage{amsmath}
\usepackage{lettrine}
\usepackage{bbding}
\usepackage{xspace}
\usepackage{textcomp}
\usepackage{textcase}
\usepackage{setspace}
\usepackage{chemformula}
\usepackage{natbib}
\usepackage{tikz}

\def\para{\ensuremath{/\kern -0.8em /}\xspace}

\def\beqn{\begin{eqnarray}}
\def\eeqn{\end{eqnarray}}
\def\beq{\begin{equation}}
\def\eeq{\end{equation}}

\newcommand{\Beq}{\begin{eqnarray*} }
\newcommand{\Eeq}{\end{eqnarray*} }

\newcommand{\sfbold}[1]{{\sffamily\bfseries #1}}

\begin{document}

\title{Beam filamentation instability drives deuterium-tritium fusion with polarized \\ neutron emission}

\author{Guanqi Qiu}
\affiliation{State  Key  Laboratory  of Nuclear  Physics  and  Technology,  Peking  University, Beijing 100871, People's Republic of China.}
\affiliation{Institute of Theoretical Physics, Chinese Academy of Sciences, Bejing 100190, China}

\author{Deji Liu}
\affiliation{State  Key  Laboratory  of Nuclear  Physics  and  Technology,  Peking  University, Beijing 100871, People's Republic of China.}

\author{Dongchi Cai}
\affiliation{State  Key  Laboratory  of Nuclear  Physics  and  Technology,  Peking  University, Beijing 100871, People's Republic of China.}
\affiliation{Institute of Theoretical Physics, Chinese Academy of Sciences, Bejing 100190, China}

\author{Ronghao Hu}
\affiliation{College of Physics, Sichuan University, Chengdu, 610065, China}

\author{Zheng Gong}
\email{Contact author: zgong92@itp.ac.cn}
\affiliation{Institute of Theoretical Physics, Chinese Academy of Sciences, Bejing 100190, China}

\author{Xueqing Yan}
\email{Contact author: x.yan@pku.edu.cn}
\affiliation{State  Key  Laboratory  of Nuclear  Physics  and  Technology,  Peking  University, Beijing 100871, People's Republic of China.}
\affiliation{Beijing Laser Acceleration Innovation Center, Huairou, Beijing, 101400, China}

\begin{abstract} 
\end{abstract}
\date{\today}
\maketitle




\maketitle

\noindent{\sfbold{Abstract}}\\
%
Filamentation instabilities are generally regarded as detrimental to fast-electron transport in fusion plasmas because they increase beam divergence and redistribute deposited energy.
Here, we use two-dimensional particle-in-cell simulations coupled to a spin-dependent deuterium-tritium fusion module to investigate whether filamentation can instead transfer energy from counterstreaming electrons to prepolarized fusion ions.
The simulations show that magnetic filaments, together with longitudinal inductive and transverse charge-separation electric fields, accelerate initially stationary deuterons and tritons to energies at which fusion reactions occur.
Over the parameter range examined, the calculated neutron yield increases with the saturated magnetic-field energy. The model further predicts anisotropic, spin-resolved neutron emission. 
The direction of maximum neutron polarization is approximately perpendicular to the dominant deuterium-tritium collision direction and evolves with the angular distribution of the reacting ions. 
These results suggest that beam filamentation can couple relativistic-electron energy to fusion ions and that polarized neutron emission may provide a reaction-weighted signature of the underlying plasma dynamics.
\\

\noindent{\sfbold{Introduction}}\\
Driven by the urgent need to overcome fossil fuel depletion and the accumulation of radioactive waste, controlled nuclear fusion stands out as a highly viable long-term energy strategy~\cite{NUCKOLLS1972,Lindl1995,Betti2016}. 
A historic milestone in this field was achieved at the National Ignition Facility (NIF) in 2022, where an input drive of $2.05\,\mathrm{MJ}$ successfully yielded $3.15\,\mathrm{MJ}$ of fusion energy~\cite{nif2024}. 
Later, NIF further pushed the fusion gain to an impressive 4.13 by 2025~\cite{nif2025}. 
Despite these breakthroughs, the conventional approach relies on megajoule-class lasers that are already operating near their maximum facility thresholds, strictly limiting future energy scaling. 
To bypass this bottleneck, fast ignition (FI) has emerged as an attractive alternative capable of functioning with significantly reduced driver energies~\cite{tabak1994,Tabak01042006,Hatchett01042006,10.1063/1.2716682}.
However, FI introduces its own complex plasma dynamics. As the relativistic electron beam traverses the core, it induces a counter-propagating return current. 
The interaction between these opposing currents triggers filamentation instabilities, fragmenting the primary beam into distinct micro-filaments~\cite{honrubia2006,karmakar2008}. Extensive three-dimensional particle-in-cell (PIC) modeling has elucidated this phenomenon. 
For instance, Honrubia et al. identified the onset of resistive beam filamentation during electron transport in ignition-scale capsules~\cite{honrubia2006}. Building on this, Karmakar et al. demonstrated that such instability-driven filamentation dramatically exacerbates beam divergence~\cite{karmakar2008}. 
This scattering effect severely degrades forward energy deposition, representing a major obstacle to maximizing yield. 
Consequently, effective suppression of beam filamentation instabilities is generally considered an important prerequisite for achieving high-gain FI.


In stark contrast to its detrimental role in laboratory settings, filamentation instabilities in astrophysical environments act as powerful mechanisms for stochastic particle acceleration, ultimately driving high-energy fusion reactions~\cite{liu2024,Silva2002}. 
When counterstreaming beams interact, the resulting instability spawns self-generated magnetic fields~\cite{silva2003,Schlickeiser_2003,Califano2006,huntington2015,Marret_2026,xu2025} and mediates the formation of collisionless shocks~\cite{Fiuza2012,Matsumoto2017,Lemoine2019}. 
This sequence cascades into stochastic turbulence, which is responsible for accelerating particles in supernova remnants~\cite{hartquist1983evidence,marcowith2006turbulence} and generating relativistic beams in gamma-ray bursts~\cite{Medvedev_1999,Piran2005,Milosavljević_2006,Ardaneh_2015}.
As these stochastically accelerated particles compress and collide, they overcome the Coulomb barrier. Such dynamics are fundamental to thermonuclear reactions powering the cosmos—ranging from hydrogen and helium burning in stellar cores~\cite{Bethe1939,burbidge1957} to heavy-element nucleosynthesis during core-collapse supernovae~\cite{WoosleyEtAl2002,Izutani_2009} and accretion-triggered Type Ia supernovae~\cite{Wang2018}. 
Drawing inspiration from these astrophysical accelerators, one can hypothesize that filamentation instability can also be deliberately harnessed to trigger fusion in laboratory experiments.

To experimentally diagnose and verify such instability-driven fusion, prepolarizing the deuterons and tritons~\cite{Rakitzis2003,PhysRevLett.118.233401,PhysRevA.73.062714} in the fuel presents a highly effective strategy. 
Polarized deuterium--tritium (DT) reactions naturally yield polarized neutrons, which are exquisitely sensitive probes extensively utilized across domains like neutron imaging~\cite{kardjilov2008,jimaging4010023,Dawson_2009}, interferometry~\cite{shahi2016}, diffraction~\cite{hayter1974,ressouche:hal-01937765,liumnte2026}, spin determination~\cite{keyworth1973,PhysRevC.18.1328}, and reflectometry~\cite{PhysRevB.52.10395,PhysRevB.64.121302,zhan2019probing,guasco2022}. 
The exceptional resolving power of polarized neutrons is well-documented: they have enabled the 3D mapping of trapped magnetic flux fields as subtle as $1.0\,\mathrm{mT}$ in superconductors measured at $7.0\,\mathrm K$~\cite{kardjilov2008}, and have proven crucial in observing nuclear--magnetic interference (corresponding to a net N'eel-vector fraction of approximately $12\%$) to manipulate altermagnetic orders in materials like MnTe~\cite{liumnte2026}. 
Furthermore, advanced reflectometry utilizing neutrons with a wavelength of $4.28\,\text{\AA}$ can rapidly map hydrogen concentrations in thin films to a precision of approximately $1\,\mathrm{at.}\%$ in under a minute, successfully extracting absolute concentrations (e.g., $93\,\mathrm{at.}\%$) when accounting for thickness variations~\cite{guasco2022}. 
Beyond solid-state physics, fusion-born neutrons intrinsically carry the thermodynamic signatures of their parent plasma.
By coupling comprehensive neutron diagnostics with theoretical models, researchers have successfully deduced core isotopic densities from neutron energy spectra and absolute yields~\cite{vingren2025}.
Consequently, polarized neutrons could potentially serve as an ideal reconstructive diagnostic tool for mapping the filamentation-instability-driven fusion process.

In this paper, we propose a paradigm shift: actively harnessing, rather than suppressing, instability to ignite fusion. 
Utilizing a newly developed spin-polarized deuterium--tritium fusion code, we consider a scenario where two counter-propagating electron beams trigger filamentation instability within a prepolarized DT fuel target. 
The resulting localized, filamentary electromagnetic fields efficiently heat the initially stationary ions. 
Since deuterons and tritons possess different masses, they experience differential acceleration under the same transient electric fields. This relative velocity drift drives highly effective ion-ion collisions, culminating in fusion reaction, with the overall reaction rate and yield dictated by the instability's growth rate and saturation amplitude. 
This scheme offers compounding advantages: prepolarizing the fuel enhances the nuclear reaction cross section~\cite{kulsrud1982,hupin2019}, while the anomalous heating provided by the filamentation instability lowers the ignition threshold, carving out a novel trajectory toward high-gain fusion. 
Furthermore, we reconstruct the fusion dynamics induced by the filamentation instability directly from the neutron signatures. 
As the emission kinematics and spin distributions of the generated polarized neutrons are exquisitely sensitive to the local electromagnetic topology, they act as an in-situ probe, allowing us to diagnose the spatiotemporal evolution of magnetic fields and plasma turbulence without requiring external probing lasers. 
Beyond advancing inertial fusion concepts, this mechanism holds significant promise for developing compact, beam- or laser-driven polarized neutron sources for advanced diagnostics. 
\\

\begin{figure*}
\centering
\includegraphics[width=0.75\linewidth]{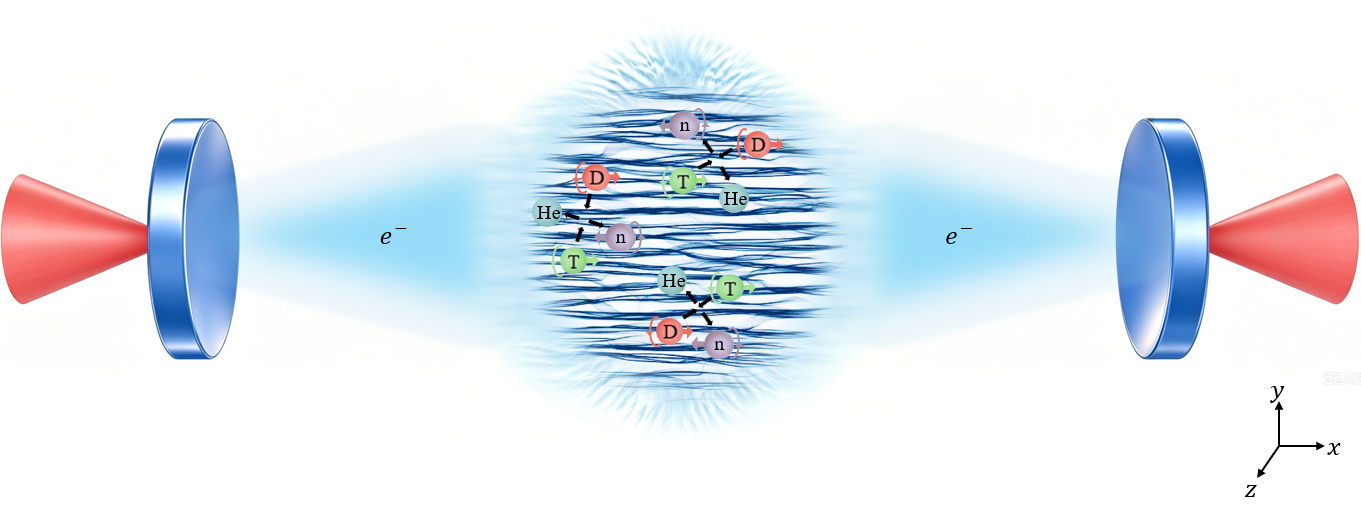}
\renewcommand{\figurename}{\sfbold{Fig.}}
\caption{Schematic. Two opposing lasers pulses irradiate targets to generate counter-propagating electron flows, which trigger current filamentation instability within the central interaction zone. This core region is pre-filled with deuteron--triton feul mixture possessing parallel spin polarization. 
As the evolving instability drives strong self-generated electromagnetic fields, the background deuterons and tritons undergo rapid heating and energetic inter-species collisions. This induces fusion reactions, culminating in the emission of polarized neutron beams.}
\label{fig:1}
\end{figure*}

\noindent{\sfbold{Results}}\\
We investigate this scheme using the two-dimensional particle-in-cell (PIC) code EPOCH~\cite{arber2015}, integrated with a custom spin-dependent fusion reaction module to capture spin polarization and deuterium--tritium dynamics. 
Within this framework, initial deuteron and triton spins are assigned values of 1 and $1/2$, respectively, while depolarization under self-consistent electromagnetic fields is governed by a spin-precession algorithm~\cite{bmt1959,PhysRevResearch.5.033115,PhysRevE.102.043215}. 
To evaluate nuclear events, the module pairs deuterons and tritons within each spatial grid cell, queries the spin-dependent differential cross section, and calculates Monte Carlo reaction probabilities based on their relative velocity. 
The polarization state of the resulting fusion neutrons is intrinsically tied to their emission kinematics, with further implementation details provided in Methods.

\begin{figure*}
\centering
\includegraphics[width=0.95\linewidth]{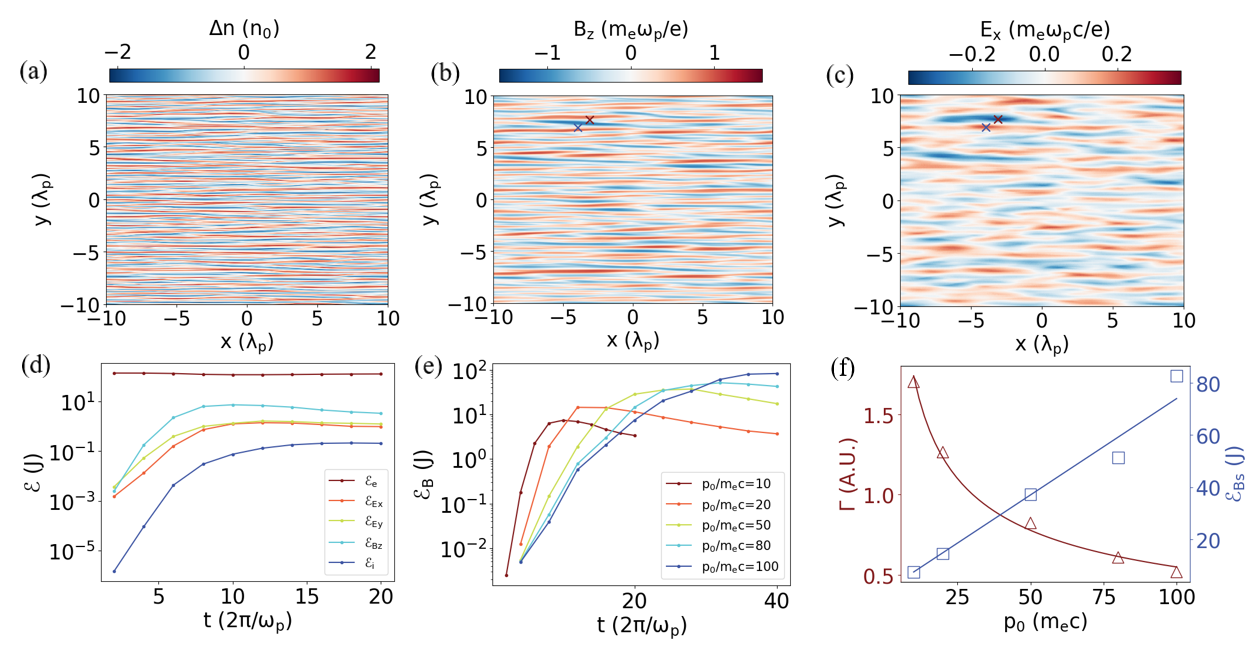}
\renewcommand{\figurename}{\sfbold{Fig.}}
\caption{Development of the filamentation instability. Distributions of (a) the density difference $\Delta n$ between right- and left-propagating electrons, (b) the self-generated magnetic field $B_z$, and (c) the longitudinal electric field $E_x$  at $t=12\pi/\omega_{p}$ for an initial electron momentum of $p_0=10m_ec$, showing clear filamentary structures. 
Here, $\lambda_p\approx10.57\,\mu\mathrm m$ is the plasma electron wavelength. 
Red and blue crosses in (b)(c) represent high-energy deuterons moving to the left and right, respectively. 
(d) Time evolution of the total electron energy $\varepsilon_e$, longitudinal electric-field energy $\varepsilon_{Ex}$, transverse electric-field energy $\varepsilon_{Ey}$, self-generated magnetic-field energy $\varepsilon_{Bz}$, and total deuteron and triton energy $\varepsilon_i$ for an initial electron momentum of $p_0=10m_ec$. 
(e) Time evolution of the magnetic-field energy for different initial electron momentum $p_0$. 
A larger initial momentum leads to slower growth of the magnetic energy but a higher saturated magnetic field strength. 
(f) Dependence of the instability growth rate $\Gamma$ and saturated magnetic energy $\varepsilon_{Bz}$ calculated from (e) on the initial electron momentum $p_0$ (markers), in agreement with the theoretical prediction (solid lines).}
\label{fig:2}
\end{figure*}

The instability is driven by two counter-propagating electron beams, each initialized with a density of $n_0=10^{19}\,\mathrm{cm}^{-3}$ (see schematic setup in Fig.~\ref{fig:1}). 
To maintain charge neutrality, the initially stationary background deuterons and tritons are uniformly distributed, each with a density of $n_0$. The initial normalized electron momentum $p_0/m_ec$ ranges from 10 to 100, where $m_e$ and $c$ denote the electron mass and speed of light, respectively.
As time progresses, the two counterpropagating electron beams interpenetrate and split into filaments, with the density of a single beam reaching up to $2.1\times10^{19}\,\mathrm{cm}^{-3}$, as shown in Fig.~\ref{fig:2}(a). This is an apparent characteristic of the plasma current filamentation instability~\cite{weibel1959}. 
Another characteristic of filamentation instability is the filamentary self-generated magnetic field $B_z$, whose strength can reach $1.1\,\mathrm{kT}$, as shown in Fig.~\ref{fig:2}(b). As the instability develops, the filaments gradually expand and merge. According to Faraday's law, a changing magnetic field produces an inductive electric field. Figure~\ref{fig:2}(c) shows the longitudinal electric field $E_x$, with a maximum field strength of $76\,\mathrm{GV/m}$. Its rate of change with time is approximately equal to the curl of the magnetic field, meaning that $E_x$ is mainly an inductive electric field generated by changes in the self-generated magnetic field of the instability. This electric field accelerates deuterons and tritons at an early stage. Figures~\ref{fig:2}(b) and (c) show the positions of high-energy deuterons moving to the left and right. 
In Fig.~\ref{fig:2}(b), the high-energy deuterons are distributed in the gaps between magnetic filaments, whereas in Fig.~\ref{fig:2}(c), these deuterons are distributed at the centres of the longitudinal electric field, with their directions of motion aligned with the electric-field direction. This shows, on the one hand, that the peak positions of the longitudinal electric field correspond to the gaps between magnetic filaments, meaning that the longitudinal electric field is mainly an inductive electric field, and, on the other hand, confirms the acceleration of deuterons by this inductive electric field. In addition, the magnetic pressure of the magnetic filaments pushes the electrons inside them outwards~\cite{Honda2000,Peterson2021}, whereas the ions remain within the magnetic filaments owing to their larger mass. This produces a transverse charge-separation field $E_y$, which also accelerates deuterons and is the source of the force responsible for the work done by the perpendicular electric field.
 
To investigate the growth of the instability, we plot the time evolution of the total electron energy $\varepsilon_e$, longitudinal electric-field energy $\varepsilon_{Ex}$, transverse electric-field energy $\varepsilon_{Ey}$, self-generated magnetic-field energy $\varepsilon_{Bz}$, and total deuteron and triton energy $\varepsilon_i$ for an initial electron-beam momentum of $p_0=10m_ec$, as shown in Fig.~\ref{fig:2}(d). 
As the instability develops, the electron energy decreases slightly, from an initial value of $141\,\mathrm J$ to $120\,\mathrm J$ at $t=20\pi/\omega_{p}$. 
Here, \(\pi\)$\approx3.14$ is the circle constant, \(\omega_{p} = \sqrt{n_{e}e^{2}/\epsilon_{0}m_{e}}\) is the plasma frequency, $n_e$ is the electron density, $e\approx1.6\times10^{-19}\,\mathrm C$ is the magnitude of the electron charge, $\epsilon_0\approx8.85\times10^{-12}\,\mathrm{F/m}$ is the vacuum permittivity, and $m_e\approx9.11\times10^{-31}\,\mathrm{kg}$ is the electron rest mass. 
This energy lost by the electrons is first transferred to the self-generated magnetic field. Subsequently, electromagnetic induction transfers part of the magnetic-field energy to the longitudinal inductive electric field, and part of the energy is transferred to the transverse electric field through the charge-separation field caused by magnetic pressure pushing electrons outwards. 
Owing to the presence of these electric fields, the initially stationary deuterons and tritons are accelerated and gain energy. 
At saturation, the total magnetic-field energy is $7.4\,\mathrm J$, the total electric-field energy is $3.0\,\mathrm J$, and the total ion energy reaches $0.21\,\mathrm J$. 
This shows that the growth of the field energy is the key to the acceleration of deuterons and tritons. 
We plot the time evolution of the magnetic-field energy for different initial electron momentum, as shown in Fig.~\ref{fig:2}(e). 
It can be found that a larger initial momentum leads to slower growth of the magnetic-field energy but a higher saturated magnetic-field strength. 
As the initial electron momentum increases from $p_0=10m_ec$ to $p_0=100m_ec$, the saturated magnetic-field energy increases from $7.4\,\mathrm J$ to $82\,\mathrm J$. 
On this basis, we plot the dependence of the growth rate of the magnetic-field energy and the saturated magnetic-field energy on the initial electron momentum, as shown by the points in Fig.~\ref{fig:2}(f). 
The filamentation instability theory predicts an instability growth rate of \(\Gamma \approx \beta\alpha\omega_p/\sqrt{\gamma}\)~\cite{Bret2004}, where $\alpha$ is the density ratio of the two electron beams, $\gamma \approx p_0/m_ec$ is the initial Lorentz factor of the electron beam, and $\beta=\sqrt{1-\gamma^{-2}}$ is the normalized electron velocity. 
Therefore, the growth rate of the filamentation instability is inversely proportional to $\sqrt{p_0}$~\cite{bret2008}. 
In addition, the saturated magnetic-field strength \(B_{s} \approx \beta c\sqrt{8\pi m_{i}n_{e}\gamma}/(1 + \beta_{th})\) is proportional to $\sqrt{\gamma}$~\cite{silva2003,Fonseca2003,Bret2010}, where $\beta_{th}$ is the thermal velocity, $n_e$ is the electron density, and $m_i$ is the ion mass. 
This indicates that the saturated magnetic-field energy is proportional to $p_0$. The solid lines show the theoretically predicted curves $\Gamma\propto p_0^{-1/2}$ and $\varepsilon_{BS}\propto p_0$, indicating that the simulation results agree with the theory. To excite a stronger instability and obtain a higher yield, we use the case with an initial electron-beam momentum of $p_0=100m_ec$ for further analysis.

\begin{figure*}
\centering
\includegraphics[width=0.95\linewidth]{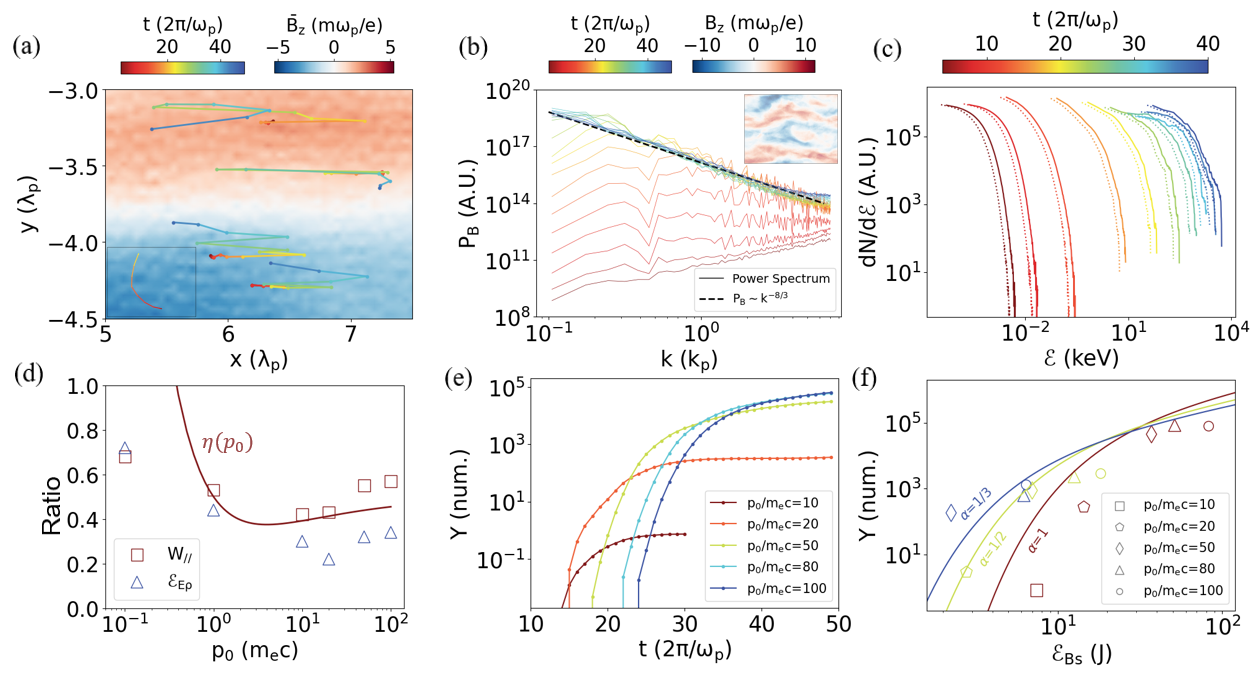}
\renewcommand{\figurename}{\sfbold{Fig.}}
\caption{Fusion reactions induced by the filamentation instability. 
(a) Distribution of the magnetic field $B_z$ averaged from $t=0$ to $t=90\pi /\omega_p$, and trajectories of some deuterons before reaction in the self-generated electromagnetic fields for an initial electron-beam momentum of $p_0=100m_ec$. The inset shows a more detailed early deuteron trajectory in a local region of $0.03\,\mu\mathrm m\times0.14\,\mu\mathrm m$, undergoing approximately semicircular motion under the magnetic field. 
(b) Time evolution of the magnetic-power spectrum  $P_B(k)$ for an initial electron momentum of $p_0=100m_ec$, converging to $P_B\sim k^{-8/3}$ after instability saturation. 
The inset shows the magnetic-field distribution at $t=90\pi/\omega_{p}$, with a turbulent structure. 
(c) Time evolution of the deuteron (solid lines) and triton (dashed lines) energy spectra for $p_0=100m_ec$. 
(d) Dependence of the ratio of parallel work to the total work done on deuterons and the ratio of the electric-field energy generated by charge separation to the total electric-field energy on the initial electron momentum. 
(e) Time evolution of the neutron yield for different initial electron momentum. A larger initial electron momentum leads to a later onset of fusion reactions but a higher yield. 
(f) Relationship between the neutron yield and the saturated magnetic-field energy for different initial electron momentum and density ratios $\alpha=n_{e1}/n_{e2}$ (markers), in agreement with theory (solid lines).}
\label{fig:3}
\end{figure*}

Considering that deuterons and tritons move over a very small region, we plot the time-averaged magnetic field in a local region and the trajectories of representative deuterons before reaction to investigate the deuteron motion in the self-generated electromagnetic fields, as shown in Fig.~\ref{fig:3}(a). 
A deuteron before reaction refers to deuteron reacting with a triton at the end of the plotted trajectory. 
Figure~\ref{fig:3}(a) shows that a deuteron is first accelerated longitudinally by the induced longitudinal electric field and then deflected by the magnetic Lorentz force, undergoing approximately semicircular motion. 
The inset of Fig.~\ref{fig:3}(a) shows the early trajectory of a deuteron under the combined action of electric and magnetic fields in a more precisely resolved region, exhibiting a clearly near-semicircular shape. 
The motion of deuterons and tritons is closely related to the evolution of the electromagnetic fields. 
We plot the time evolution of the magnetic-power spectrum $P_B(k)$, as shown in Fig.~\ref{fig:3}(b). 
After the instability saturates, the magnetic-power spectrum is $P_B\propto k^{-8/3}$, where $k$ is the wavenumber of the filamentation instability, consistent with the compressible turbulence spectra actually observed in space environments such as the magnetosheath and solar wind~\cite{chefranov2021,boldyrev2012}. 
This means that the acceleration of deuterons and tritons may be similar to acceleration in astrophysical turbulence. The inset shows the magnetic-field distribution after instability saturation at $t=90\pi/\omega_{p}$, which has a turbulence-like appearance. 
The continuous variations in these electromagnetic fields cause the deuteron trajectories in Fig.~\ref{fig:3}(a) to resemble trajectories of stochastic acceleration in turbulence.
Deuterons and tritons are accelerated by electromagnetic fields, and the time evolution of their energy spectra is shown in Fig.~\ref{fig:3}(c). 
At the instability saturation time of $t=80\pi/\omega_{p}$, the deuteron and triton energies have almost stopped accelerating. 
At this time, the maximum deuteron energy reaches $5212\,\mathrm{keV}$ and the maximum triton energy reaches $3549\,\mathrm{keV}$, approximately two-thirds of the deuteron energy.

Considering the effects of different initial electron momentum on the electric field and on the work done by the electric field on deuterons, we plot the dependence of the ratio of parallel work to the total work done on deuterons and the ratio of electric-field energy generated by charge separation to the total electric-field energy on the initial electron momentum. Both show a decrease followed by an increase. As $p_0$ increases, the parallel-work ratio decreases from $68\%$ to $42\%$ and then rises to $57\%$, while the inductive-electric-field energy ratio decreases from $72\%$ to $22\%$ and then rises to $34\%$. The electric-field energy generated by charge separation is obtained by substituting the charge density into Poisson's equation. A lower charge-separation-field ratio means a higher inductive-electric-field ratio, which accelerates deuterons longitudinally. The magnetic field quickly deflects deuterons into transverse motion, so the ratio of transverse work becomes higher and the parallel-work ratio becomes lower. This shows that the trends of the charge-separation-field ratio and the parallel-work ratio are consistent. Consider a deuteron moving from the bottom of a circle along the right semicircle towards the top and undergoing a reaction along the way. A larger initial electron momentum leads to a lower instability growth rate, a longer growth time and a larger angle of rotation. If the reaction occurs before one-quarter of a revolution, the perpendicular-work ratio continues to increase. If the reaction occurs between one-quarter and one-half of a revolution, the perpendicular-work ratio continues to decrease. Therefore, as the initial electron momentum increases, the perpendicular-work ratio first increases and then decreases, whereas the parallel-work ratio first decreases and then increases. We also propose a theoretical model to describe this variation, in which the parallel-work ratio or the charge-separation-field energy ratio satisfies \(\eta(p_0)\sim 1 - 1/\sqrt{p_{0}} + 1/p_{0}\), where $p_0$ is the initial electron momentum. The detailed derivation is given in Methods. The theoretical curve \(\eta(p_0)\) is shown as a solid line in Fig.~\ref{fig:3}(d) and agrees with the trend of the simulation points. The larger discrepancy at lower initial electron momentum may be due to neglecting the development of the two-stream and oblique instabilities.

\begin{figure*}
\centering
\includegraphics[width=0.75\linewidth]{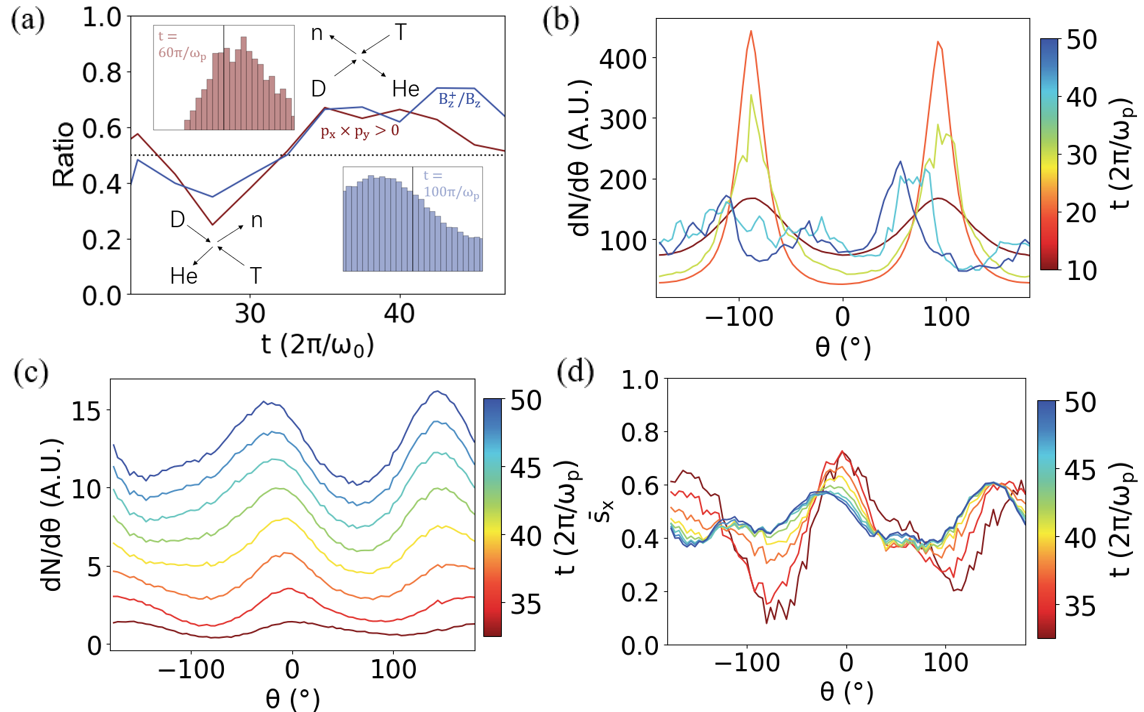}
\renewcommand{\figurename}{\sfbold{Fig.}}
\caption{Generation of polarized neutrons. For an initial electron momentum of $p_0=100m_ec$: (a) Time evolution of the ratio of deuterons to be reacted satisfying \(p_{x} \times p_{y} > 0\) and the ratio of the magnetic field along the positive $z$ direction experienced by the reacting deuterons. The upper-left inset shows the energy--angle distribution of neutrons with $s_x=-\hbar/2$ near $0^\circ$ (black line) at $t=60\pi/\omega_{p}$, with a maximum neutron emission angle of $16^\circ$. The lower-right inset shows the energy--angle distribution of neutrons with $s_x=-\hbar/2$ near $0^\circ$ at $t=100\pi/\omega_{p}$, with a maximum neutron emission angle of $-28^\circ$. (b) Time evolution of the deuteron angular distribution. (c) Time evolution of the angular distribution of neutrons with \(\text{$s_x=-\hbar/2$}\). (d) Time evolution of the angular distribution of neutron polarization. The deuteron angle shifts from $92^\circ$ towards $56^\circ$, causing the angular peak of neutrons with \(\text{$s_x=-\hbar/2$}\) to shift from $0^\circ$ towards $-28^\circ$.}
\label{fig:4}
\end{figure*}

The above analysis uses deuterons as a representative example. Tritons undergo the same acceleration process but have lower velocities owing to their larger mass. The accelerated deuterons and tritons collide, causing fusion reactions and producing polarized neutrons. The time evolution of the neutron yield $Y$ for different initial electron momentum is shown in Fig.~\ref{fig:3}(e). For $p_0=100m_ec$, the neutron yield is $6.4\times10^4$, nearly five orders of magnitude higher than the yield of $0.74$ for $p_0=10m_ec$. The figure also shows that a larger initial electron momentum leads to a later onset of fusion reactions but a higher yield. This is because a larger initial electron momentum leads to a lower instability growth rate, so it takes longer to grow to the strength required to induce fusion reactions, and the fusion reactions therefore start later. However, a larger initial electron momentum means that more energy is transferred from the electrons to the electromagnetic fields, which in turn accelerate deuterons and tritons to higher velocities. The increase in yield from $p_0=10m_ec$ to $p_0=50m_ec$ is mainly due to the increase in the cross section caused by the higher deuteron and triton velocities. When $p_0>50m_ec$, however, the yield no longer increases markedly. This is because the relative velocity of the deuterons and tritons exceeds the velocity corresponding to the maximum cross section, so the cross section decreases as the velocity increases~\cite{bosch1992}. However, a higher velocity means a higher collision frequency. As a result, there is only a small increase in yield with the initial electron momentum.

In addition, to illustrate asymmetric incidence of the electron beams, we plot the relationship between the neutron yield and the saturated magnetic-field energy for different electron-beam density ratios $\alpha=n_{e1}/n_{e2}$, as shown by the points in Fig.~\ref{fig:3}f. Here, $n_{e1}$ is the density of the right-propagating electron beam and $n_{e2}$ is the density of the left-propagating electron beam. The figure shows that the points at the same initial electron momentum but different $\alpha$ values approximately follow a linear relationship, indicating that the yield depends only on the saturated magnetic-field energy. However, a smaller $\alpha$ results in weaker filamentation instability and a lower saturated magnetic-field energy at the same initial electron momentum. This leads to lower deuteron and triton energies and, consequently, a lower yield. For $p_0=100m_ec$, the neutron yield is $2.8\times10^3$ at $\alpha=1/2$, a reduction of $96\%$ compared with $\alpha=1$, and $1.3\times10^3$ at $\alpha=1/3$, a reduction of $98\%$ compared with $\alpha=1$. We estimate the yield $Y$ using the number of deuterons for which the cross section is greater than $0.1\,\mathrm b$, that is, \(Y\sim\int_{\varepsilon_{1}}^{\varepsilon_{2}}{e^{- \varepsilon/\alpha\gamma}dE} = \gamma\alpha(e_{N}^{- \frac{\varepsilon_{1}}{\alpha\gamma}} - e_{N}^{- \frac{\varepsilon_{2}}{\alpha\gamma}})\), where \(\varepsilon_{1},\varepsilon_{2}\) are the deuteron energies corresponding to a cross section of $0.1\,\mathrm b$, $\gamma \approx p_0/m_ec$ is the initial electron Lorentz factor and $e_N\approx2.71$ is the base of the natural logarithm. The detailed derivation is given in Methods. The estimates are shown as solid lines in Fig.~\ref{fig:3}(f), which are in agreement with the simulation results.

Polarized deuterium--tritium reactions can produce polarized neutrons. We found negligible depolarization of deuterons and tritons as a result of their heavy mass and low velocity. Because neutrons with spins antiparallel to those of the deuterons and tritons are fully polarized in the direction perpendicular to the deuteron--triton collision direction, the deuteron--triton collision angle can be reconstructed from the angular distribution of neutron polarization. Figure~\ref{fig:4}(a) shows the time evolution of the ratio of reacting deuterons for which the product of the longitudinal and transverse momentum satisfies \(p_{x} \times p_{y} > 0\), and the ratio of the field $B_z^+$ along the positive $z$ direction to the total $B_z$ magnetic-field energy experienced by the deuterons to be reacted, for an initial electron-beam momentum of $p_0=100m_ec$. Here, deuterons to be reacted refer to deuterons that have already reacted by the next saved output. When a greater ratio of the magnetic field points along the negative $z$ direction, deuterons and tritons tend to move along the diagonals of the second and fourth quadrants after approximately semicircular motion, which is $p_{x} \times p_{y} < 0$. Because the direction with the largest number of neutrons with longitudinal spin $s_x=-\hbar/2$ is perpendicular to the deuteron--triton collision direction, the neutrons with $s_x=-\hbar/2$ then tend to be emitted along the diagonals of the first and third quadrants, giving a neutron emission angle of $16^\circ$ at $t=60\pi/\omega_{p}$ in the inset. Conversely, when a greater ratio of the magnetic field points along the positive $z$ direction, deuterons and tritons tend to move along the diagonals of the first and third quadrants, which is $p_{x} \times p_{y} > 0$. At this time, neutrons with \(\text{$s_x=-\hbar/2$}\) tend to be emitted along the diagonals of the second and fourth quadrants, giving a neutron emission angle of $-28^\circ$ at $t=100\pi/\omega_{p}$ in the inset.

Figure~\ref{fig:4}(b) shows the time evolution of the deuteron angular distribution. As the instability develops, the deuteron angle shifts from $92^\circ$ to $56^\circ$, causing the angular peak of neutrons with \(\text{$s_x=-\hbar/2$}\) to shift from $0^\circ$ to $-28^\circ$, as shown in Fig.~\ref{fig:4}(c). This is because the direction with the largest number of neutrons with $s_x=-\hbar/2$ is perpendicular to the collision direction, and the collision direction is approximately along the deuteron direction of motion. That is, the direction with the largest number of neutrons with $s_x=-\hbar/2$ is approximately perpendicular to the deuteron direction of motion, so a decrease in the deuteron peak angle also leads to a decrease in the neutron peak angle. Because the yield of neutrons with $s_x=\hbar/2$ is lower, the neutron polarization mainly depends on the number of neutrons with $s_x=-\hbar/2$. Therefore, the peak in the number of neutrons with $s_x=-\hbar/2$ approximately corresponds to the peak in polarization, and the polarization peak shifts from $-4^\circ$ to $-20^\circ$, as shown in Fig.~\ref{fig:4}(d). In addition to indicating that deuterons and tritons mainly collide and react in a nearly transverse direction, this also means that a polarized neutron beam with a polarization of approximately $60\%$ can be obtained in the $-20^\circ$ direction for applications such as probing.
\\

\noindent{\sfbold{Conclusion}}\\
In summary, utilizing a newly developed fusion reaction code incorporating spin, we use two counterstreaming electron beams to trigger polarized deuterium--tritium fusion and show the potential of producing polarized neutron beams. The two counterstreaming electron beams excite filamentation instability, generating self-generated magnetic fields and further exciting self-generated electric fields, which accelerate prepolarized deuterons and tritons. 
Their collisions trigger fusion reactions and produce polarized neutron beams.
This study turns instability from a mechanism conventionally considered to require suppression in fusion into a factor that can induce fusion reactions, providing an alternative promising way to initiate fusion reactions. The magnetic-power spectrum after instability saturation agrees with that in the solar wind, and deuterons and tritons collide and react in turbulence-like structures, potentially offering a possible explanation for the formation of the solar-wind magnetosheath and the synthesis of heavy elements near high-energy cosmic rays in the universe. 
The angular characteristics of generated polarized neutrons might also offer a possibility to help retrieve and understand the in-situ spin-polarized D-T nuclear reaction associated with filamentation dynamics.
\\

\noindent{\sfbold{Methods}}\\
\noindent{\sfbold{1. Fusion reaction code incorporating spin polarization}}\\
\noindent{\sfbold{1.1 Polarized reaction theory}}\\
The deuteron spin is $\hbar$ and the triton spin is $\hbar/2$, so the total angular momentum $J$ after deuterium--tritium fusion can take the values $\hbar/2$ or $2\hbar/3$. The deuteron and triton wave functions are, respectively,
\begin{equation}
\left|\chi_D\right\rangle=\sum_{m_2}a_{m_2}\left|m_2\right\rangle,
\qquad
\left|\chi_T\right\rangle=\sum_{m_3}b_{m_3}\left|m_3\right\rangle,
\label{eq:1}
\end{equation}
where $m_2$ and $m_3$ are the spin angular momentum quantum numbers of the deuteron and triton, respectively, and $a_{m_2}$ and $b_{m_3}$ are the coefficients. During fusion, the wave function of the fused nucleus is
\begin{equation}
\left|\chi\right\rangle
=\left|\chi_D\right\rangle\left|\chi_T\right\rangle
=\sum_{m_2,m_3}a_{m_2}b_{m_3}\left|m_2\,m_3\right\rangle.
\label{eq:2}
\end{equation}
The reaction cross section is
\begin{equation}
\begin{aligned}
\sigma
&=\sum_{J=1/2,\,3/2}\sum_{m=-J}^{J}
\sigma_0\left|\left\langle J,m\mid\chi\right\rangle\right|^2\\
&=\sum_{J=1/2,\,3/2}\sum_{m_2,m_3}
d_{m_2}t_{m_3}
\left|\left\langle J,m_2+m_3\mid m_2,m_3\right\rangle\right|^2\sigma_0,
\end{aligned}
\label{eq:3}
\end{equation}
where $d_{m_2}$ and $t_{m_3}$ are coefficients, $\sigma_0$ is the cross section without polarization. According to the Clebsch--Gordan coefficient table~\cite{condon1935theory},
\[
|\frac{3}{2}\ \frac{3}{2} > = |1 > |\frac{1}{2} >,
\]
\[
|\frac{3}{2}\ \frac{1}{2} > = \sqrt{\frac{1}{3}}\ |1 > | - \frac{1}{2} > + \sqrt{\frac{2}{3}}|0 > |\frac{1}{2} >,
\]
\[
|\frac{3}{2}\  - \frac{1}{2} > = \sqrt{\frac{2}{3}}|0 > | - \frac{1}{2} > + \sqrt{\frac{1}{3}}| - 1 > |\frac{1}{2} >,
\]
\[
|\frac{3}{2}\  - \frac{3}{2} > = | - 1 > | - \frac{1}{2} >,
\]
\[
|\frac{1}{2}\ \frac{1}{2} > = \sqrt{\frac{2}{3}}\ |1 > | - \frac{1}{2} > - \sqrt{\frac{1}{3}}|0 > |\frac{1}{2} >,
\]
\[
|\frac{1}{2}\  - \frac{1}{2} > = \sqrt{\frac{1}{3}}|0 > | - \frac{1}{2} > - \sqrt{\frac{2}{3}}| - 1 > |\frac{1}{2} >.
\]
Therefore, the cross section can be simplified to~\cite{kulsrud1986}
\begin{equation}
\sigma=\left(a+\frac23b+\frac13c\right)\sigma_{3/2}
+\left(\frac13b+\frac23c\right)\sigma_{1/2},
\label{eq:4}
\end{equation}
where $a=d_1t_{1/2}+d_{-1}t_{-1/2},~b=d_0(t_{1/2}+t_{-1/2})=d_0,~c=d_1t_{-1/2}+d_{-1}t_{1/2}$, $\sigma_{3/2}$ is the cross section for $J=3/2$ and $\sigma_{1/2}$ is the cross section for $J=1/2$. Because $\sigma_{3/2}\gg\sigma_{1/2}$, $\sigma_{1/2}$ can be neglected. When the deuterons and tritons are initially unpolarized, all $d$ and $t$ are $1/2$, which means $a=b=c=1/3$, so $\sigma_0=\tfrac23\sigma_{3/2}$, or $\sigma_{3/2}=\tfrac32\sigma_0$. When the deuterons and tritons are initially polarized parallel to one another, which means $a=1$, $b=0$ and $c=0$, $\sigma_a=\sigma_{3/2}=\tfrac32\sigma_0$. This is why parallel polarization increases the deuterium--tritium fusion cross section by $50\%$~\cite{kulsrud1982,hupin2019,Reichwein_2025}. When the deuterons and tritons are initially polarized antiparallel to one another, which is $a=0$, $b=0$ and $c=1$, $\sigma_a=\tfrac13\sigma_{3/2}=\tfrac12\sigma_0$, meaning that antiparallel polarization reduces the cross section by $50\%$.
The fused nucleus subsequently splits into a He and a neutron. Because the fused nucleus has even parity, the He nucleus has spin 0 and the neutron has spin $\hbar/2$, the neutron wave function can be obtained from angular-momentum conservation as
\begin{equation}
\begin{aligned}
\left|\psi_f^\ell\right\rangle
&=\left|\ell,m_\ell m_s\right\rangle
\left\langle\ell,m_\ell m_s\mid JM\right\rangle
\left(\sigma_0d_{m_2}t_{m_3}\right)^{1/2}\\
&\quad\times\left\langle JM\mid m_2,m_3\right\rangle
\end{aligned}
\label{eq:5}
\end{equation}
where $\ell=2$ is the orbital angular momentum of the neutron, $m_l$ is the orbital angular momentum quantum number of the neutron, $m_s$ is the spin angular momentum quantum number of the neutron. According to the Clebsch--Gordan coefficient table~\cite{condon1935theory},
\[
|\frac{3}{2}\ \frac{3}{2} > = \frac{2}{\sqrt{5}}\ |2 > | - \frac{1}{2} > - \frac{1}{\sqrt{5}}|1 > |\frac{1}{2} >
\]
\[
|\frac{3}{2}\ \frac{1}{2} > = \sqrt{\frac{3}{5}}\ |1 > | - \frac{1}{2} > - \sqrt{\frac{2}{5}}|0 > |\frac{1}{2} >
\]
\[
|\frac{3}{2}\  - \frac{1}{2} > = \sqrt{\frac{2}{5}}|0 > | - \frac{1}{2} > - \sqrt{\frac{3}{5}}| - 1 > |\frac{1}{2} >
\]
\[
|\frac{3}{2}\  - \frac{3}{2} > = \frac{1}{\sqrt{5}}\ | - 1 > | - \frac{1}{2} > - \frac{2}{\sqrt{5}}| - 2 > |\frac{1}{2} >
\]
\[
|\frac{3}{2}\ \frac{3}{2} > = |1 > |\frac{1}{2} >
\]
\[
|\frac{3}{2}\ \frac{1}{2} > = \sqrt{\frac{1}{3}}\ |1 > | - \frac{1}{2} > + \sqrt{\frac{2}{3}}|0 > |\frac{1}{2} >
\]
\[
|\frac{3}{2}\  - \frac{1}{2} > = \sqrt{\frac{2}{3}}|0 > | - \frac{1}{2} > + \sqrt{\frac{1}{3}}| - 1 > |\frac{1}{2} >
\]
\[
|\frac{3}{2}\  - \frac{3}{2} > = | - 1 > | - \frac{1}{2} >
\]

Decomposing the total-angular-momentum state $J=3/2$ of the fused nucleus gives the neutron wave function as
\begin{equation}
\begin{aligned}
&
\psi_{n} = \left( \frac{\sigma_{3/2}}{5} \right)^{\frac{1}{2}}\{| - \frac{1}{2} > \bigl\lbrack 2Y_{22}\left( d_{1}t_{1/2} \right)^{1/2} \\
&\quad + Y_{21}\left( d_{1}t_{- 1/2} \right)^{1/2} + {2Y}_{20}\left( d_{0}t_{- 1/2}/3 \right)^{1/2} \\
&\quad + Y_{21}\left( 2d_{0}t_{1/2} \right)^{1/2} + Y_{20}\left( 2d_{- 1}t_{1/2}/3 \right)^{1/2} \\
&\quad + Y_{2 - 1}\left( d_{- 1}t_{- 1/2} \right)^{1/2} \\
&- |\frac{1}{2} > \lbrack Y_{21}\left( d_{1}t_{1/2} \right)^{1/2} \\
&\quad + Y_{20}\left( 2d_{1}t_{- 1/2}/3 \right)^{1/2} + Y_{2 - 1}\left( 2d_{0}t_{- 1/2} \right)^{1/2} \\
&\quad + {2Y}_{20}\left( d_{0}t_{1/2}/3 \right)^{1/2} + Y_{2 - 1}\left( d_{- 1}t_{1/2} \right)^{1/2} \\
&\quad + {2Y}_{2 - 2}\left( d_{- 1}t_{- 1/2} \right)^{1/2}\rbrack\}
\end{aligned}
\label{eq:6}
\end{equation}

Here, $Y_{\ell m}$ are spherical harmonics, $\ell$ is the angular-momentum quantum number, and $m$ is the magnetic quantum number. Their expressions for $\ell=2$ are as follows:
\[
\begin{aligned}
Y_2^{-2}(\theta,\varphi)
&=\frac14\sqrt{\frac{15}{2\pi}}e^{-2i\varphi}\sin^2\theta
=\frac14\sqrt{\frac{15}{2\pi}}\frac{(x-iy)^2}{r^2}\\[4pt]
Y_2^{-1}(\theta,\varphi)
&=\frac12\sqrt{\frac{15}{2\pi}}e^{-i\varphi}\sin\theta\cdot\cos\theta
=\frac12\sqrt{\frac{15}{2\pi}}\frac{(x-iy)z}{r^2}\\[4pt]
Y_2^0(\theta,\varphi)
&=\frac14\sqrt{\frac5\pi}(3\cos^2\theta-1)
=\frac14\sqrt{\frac5\pi}\frac{2z^2-x^2-y^2}{r^2}\\[4pt]
Y_2^1(\theta,\varphi)
&=-\frac12\sqrt{\frac{15}{2\pi}}e^{i\varphi}\sin\theta\cdot\cos\theta
=-\frac12\sqrt{\frac{15}{2\pi}}\frac{(x+iy)z}{r^2}\\[4pt]
Y_2^2(\theta,\varphi)
&=\frac14\sqrt{\frac{15}{2\pi}}e^{2i\varphi}\sin^2\theta
=\frac14\sqrt{\frac{15}{2\pi}}\frac{(x+iy)^2}{r^2}
\end{aligned}
\]
When the initial deuteron spin is $s_x=\hbar$ and the initial triton spin is $s_x=\hbar/2$, $d_1=1$, $t_{1/2}=1$ and $d_0=d_{-1}=t_{-1/2}=0$. Squaring the wave function gives the neutron angular distribution as

\begin{equation}
\begin{aligned}
\frac{d\sigma}{d\Omega}
&=\sum_{m_s=\pm(1/2)}\left|\psi_f\right|^2\\
&=\frac1{2\pi}\biggl[\sigma_{3/2}\biggl[\frac34\sin^2\theta\,a
+\frac14(1+3\cos^2\theta)\left(\frac23b+\frac13c\right)\biggr]\\
&\qquad\quad+\sigma_{1/2}\left(\frac13c+\frac16b\right)\biggr]
\end{aligned}
\label{eq:7}
\end{equation}
At $90^\circ$, $|Y_{22}|^2=15/32\pi$ and $|Y_{21}|^2=0$, so all neutrons in this direction have spin $s_x=-\hbar/2$. 
\\

\begin{figure}
\centering
\includegraphics[width=0.75\linewidth]{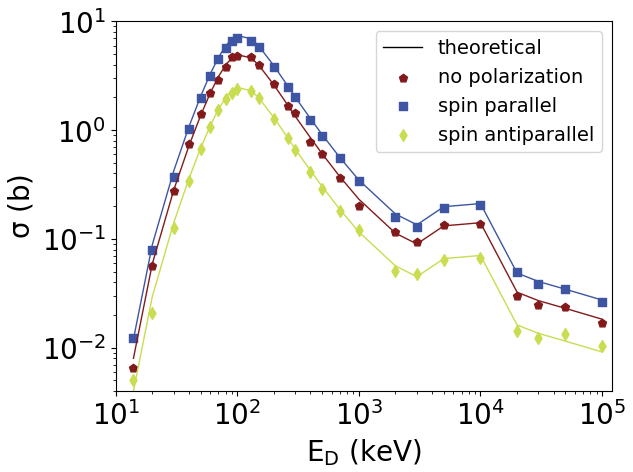}
\renewcommand{\figurename}{\sfbold{Fig.}}
\caption{Benchmark of the fusion reaction code, taking into account the effect of spin on the reaction cross section. Points show the cross sections calculated by the code, and solid lines show the theoretical cross sections.}
\label{fig:S1}
\end{figure}

\noindent{\sfbold{1.2 Fusion code}}\\
Suppose that at a certain time there are $N_1$ deuterons and $N_2$ tritons at a cell grid. Each deuteron and triton is randomly paired~\cite{higginson2019,HIGGINSON2020109450}, giving $N=\min(N_1,N_2)$ particle pairs. For convenience in the calculation, the kinematic parameters of the deuterons and tritons are transformed to the centre-of-momentum frame. The reaction probability is calculated as $P=NW\sigma v\Delta t/V$, where $v$ is the relative velocity of the deuteron and triton in a particle pair, $W$ is the weight, $\Delta t$ is the time step and $V$ is the cell volume. A random number between 0 and 1 is then generated, and the reaction occurs if it is smaller than $P$. According to momentum and energy conservation, the momentum and angles of the products can be calculated, with the neutron emission direction determined by the neutron wave function \(\psi_{n}\). Finally, the products are generated, the reactants are removed, and the kinematic parameters of the neutron and He nucleus are transformed back to the laboratory frame. When calculating the reaction probability, the cross section is obtained by table lookup: below the minimum energy, the cross section is zero; within the table range, linear interpolation is used; above the table range, linear extrapolation is used. The relationship between the reaction cross section and the deuteron and triton spins is given in Eq.~\ref{eq:4}. The code benchmark is shown in Fig.~\ref{fig:S1}, where the cross section obtained from the yield calculated by the code agrees with the theoretical cross section.
\\

\begin{figure}
\centering
\includegraphics[width=0.75\linewidth]{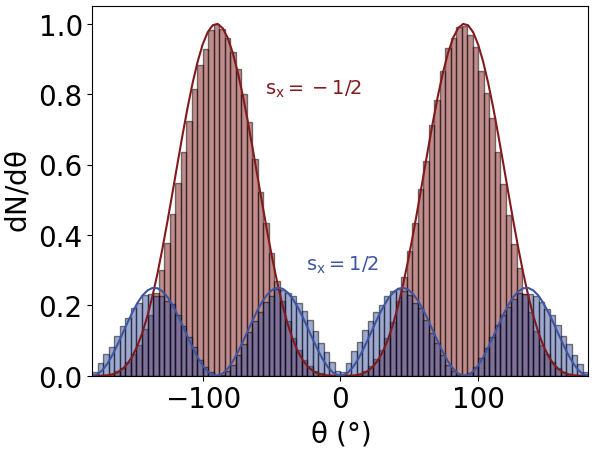}
\renewcommand{\figurename}{\sfbold{Fig.}}
\caption{Angular distributions of neutrons with spins $s_x=\hbar/2$ and $s_x=-\hbar/2$ calculated by the code for a deuteron beam incident normally on a tritium target. Solid lines show the theoretical results.}
\label{fig:S2}
\end{figure}

\noindent{\sfbold{1.3 Spin module}}\\
To describe the effect of prepolarized deuterium and tritium on fusion reactions, we added the change in spin due to precession and the effect of spin on fusion reactions to the code. The initial deuteron spin is $s_x=\hbar$, $s_y=0$, $s_z=0$, and the initial triton spin is $s_x=\hbar/2$, $s_y=0$, $s_z=0$. The change in deuteron and triton spins caused by the self-generated electromagnetic fields in the plasma is

\begin{equation}
\begin{aligned}
\frac{ds}{dt}&=s\times\Omega,\\
\Omega&=\frac{e}{2m_i}\biggl\{
g\left(\frac{B}{\gamma_i}-\frac1{\gamma_i+1}\frac{v}{c^2}\times E\right)\\
&\qquad\quad-(g-2)\frac{\gamma_i}{\gamma_i+1}\frac{v}{c^2}\times(E+v\times B)
\biggr\},
\end{aligned}
\label{eq:8}
\end{equation}
where $m_i$ is the mass of a deuteron or a triton, $g=1.714$ for deuterons or $g=17.836$ for tritons is the Lande $g$ factor, $\gamma_i$ is the Lorentz factor of a deuteron or a triton, $B$ is the magnetic filed and $E$ is the electric field. During a fusion reaction, the spin component $s_x$ of each deuteron--triton pair is read. For a deuteron, if $s_x>1/3$, it is assumed that $d_1=1$, $d_0=0$ and $d_{-1}=0$; if $s_x<-1/3$, it is assumed that $d_1=0$, $d_0=0$ and $d_{-1}=1$; and if $-1/3\leq s_x\leq1/3$, it is assumed that $d_1=0$, $d_0=1$ and $d_{-1}=0$. For a triton, if $s_x>0$, it is assumed that $t_{1/2}=1$ and $t_{-1/2}=0$ in the neutron wave function \(\psi_{n}\); if $s_x\leq0$, then $t_{1/2}=0$ and $t_{-1/2}=1$. At $90^\circ$, the coefficient of the neutron $1/2$-spin wave function $|1/2\rangle$ is zero, so all neutrons emitted in this direction have spin $-1/2$. The code benchmark is shown in Fig.~\ref{fig:S2}, where the angular distribution calculated by the code (histogram) agrees with the theoretical expectation (solid lines).
\\

\noindent{\sfbold{2. Simulation setup}}\\
The simulations use two infinitely extended electron beams, each with a density of $n_0=10^{19}\,\mathrm{cm}^{-3}$, propagating along the positive and negative $x$ directions, respectively. The initial electron-beam momentum $p_0/m_ec$ ranges from 10 to 100, where $m_e$ is the electron mass and $c$ is the speed of light. To maintain charge neutrality, prepolarized deuterons are stationary in the simulation region, with densities of $n_0$. The simulation region is a square of $20\lambda_p\times20\lambda_p$ with periodic boundary conditions, where $\lambda_p\approx10.57\,\mu\mathrm m$ is the plasma wavelength. When reading physical quantities such as energy and yield, the third dimension is estimated as $20\lambda_p$.
\\

\noindent{\sfbold{3. Theoretical model}}\\
\noindent{\sfbold{3.1 Parallel-work ratio}}\\
The growth rate of filamentation instability satisfies \(\Gamma\sim\frac{\omega_p}{\sqrt{\gamma}}\)~\cite{Bret2004}, where $\gamma \approx p_0/m_ec$ is the initial electron Lorentz factor. Deuterons and tritons move over a very small region, and the electromagnetic fields they experience can be approximated as spatially invariant, which is \(e^{i(kx - \omega t)} \approx 1\), where $k$ is the wavenumber and $\omega$ is the laser angular frequency. The simulations show that $E_x$ and $E_y$ are of the same order of magnitude, so we assume \(\frac{E_{y}^{2}}{E_{x}^{2}}\sim 1\), with \(\frac{cB_{0}}{E_{0}} = \frac{ck}{\Gamma}\sim 1\), where $k$ is the instability wavenumber, $B_0$ is the initial magnetic field and $E_0$ is the initial electric field. Assuming \(E_{x} = E_{0}e^{\frac{\omega_pt}{\sqrt{\gamma}}},\ E_{y} = - E_{0}e^{\frac{\omega_pt}{\sqrt{\gamma}}},\ B_{z} = B_{0}e^{\frac{\omega_pt}{\sqrt{\gamma}}}\), the deuteron equations of motion are
\begin{equation}
\left\{ \begin{array}{r}
m_{D}\dot{v_{x}} = - eE_{x} - ev_{y}B_{z}\text{~} \\
m_{D}\dot{v_{y}} = - eE_{y} + ev_{x}B_{z}
\end{array} \right.\
\label{eq:9}
\end{equation}
where $m_D$ is the mass of a deuteron. Solving gives the longitudinal velocity
\begin{equation}
v_{x} = \cos\Theta - 1 - \sin\Theta
\label{eq:10}
\end{equation}
where \(\Theta = \frac{eB_{0}e^{\frac{\omega_pt}{\sqrt{\gamma}}} - eB_{0}}{m_{D}k}\sim\frac{B_{sat}/B_{0} - 1}{1/\sqrt{\gamma}}\sim\gamma - \sqrt{\gamma}\), and \(B_{sat}\) is the saturated magnetic-field strength. Therefore, the parallel work is
\begin{equation}
\begin{aligned}
W_{x} &= - e\int E_{x}v_{x}dt \\
&= m_{D}c^2\Theta - m_{D}c^2\sin\Theta + m_{D}c^2 - m_{D}c^2\cos\Theta
\end{aligned}
\label{eq:11}
\end{equation}
Taking a period average gives
\begin{equation}
\overline{W_{x}}\sim(\gamma - \sqrt{\gamma} + 1) m_{D}c^2.
\label{eq:12}
\end{equation}
Considering that the total work is \(W_\mathrm{tot}\sim\gamma  m_{D}c^2\), the parallel-work ratio is
\begin{equation}
\eta = \frac{W_{x}}{W_\mathrm{tot}}\ \sim 1 - \frac{1}{\sqrt{\gamma}} + \frac{1}{\gamma}.
\label{eq:13}
\end{equation}

\bigskip
\noindent{\sfbold{3.2 Charge-separation-field energy ratio}}\\

According to Faraday's law,
\begin{equation}
\frac{\partial B}{\partial t} = \frac{\omega_p}{\sqrt{\gamma}}B = - \frac{\partial E_{in}}{\partial y} \approx \frac{E_{in}}{\lambda} = \frac{\omega_p E_{in}}{c\sqrt{\gamma}}
\label{eq:14}
\end{equation}
where $E_{in}$ is the inductive electric field and $\lambda$ is the instability wavelength, so \(B\sim E_{in}\). The perpendicular work is
\begin{equation}
W_{\bot}\propto\int\frac{\partial B^{2}}{\partial y}dy = B^{2}
\label{eq:15}
\end{equation}
Therefore, \(W_{\bot} \propto E_{in}^{2}\sim c^2B^{2}\). Thus, the perpendicular-work ratio is approximately equal to the ratio of the inductive-electric-field energy to the total electric-field energy \(E_{tot}^{2}\),
\begin{equation}
W_{\bot}/W_\mathrm{tot}\sim E_{in}^{2}/E_{tot}^{2}.
\label{eq:16}
\end{equation}
that is, the parallel-work ratio is approximately equal to the charge-separation-field energy ratio.

\bigskip
\noindent{\sfbold{3.3 Yield estimate}}\\

The total work done by the electric field on deuterons is \(W_\mathrm{tot}\sim\alpha\gamma m_{D}c^2\), where $\alpha=n_{e1}/n_{e2}$ is the density ratio of the two electron beams. Deuterons follow a Boltzmann distribution \(N\sim e_N^{- \varepsilon/k_{B}T_{i}}\), where $k_B\approx1.38\times10^{-23}\,\mathrm{J/K}$ is the Boltzmann constant, $T_i$ is the deuteron temperature, and \(\varepsilon\) is the deuteron energy. The total work done by the electric field on the deuterons is equal to the sum of the deuteron energies,

\begin{equation}
W_\mathrm{tot}\sim\int e^{- \frac{\varepsilon}{k_{B}T_{i}}}d\varepsilon = k_{B}T_{i}
\label{eq:17}
\end{equation}
which is \(k_{B}T_{i}\sim\alpha\gamma\). The yield is estimated using the number of deuterons for which the cross section is greater than $0.1\,\mathrm b$,
\begin{equation}
Y\sim\int_{\varepsilon_{1}}^{\varepsilon_{2}}{e^{- \varepsilon/k_{B}T_{i}}d\varepsilon} = \alpha\gamma(e^{- \frac{\varepsilon_{1}}{\alpha\gamma}} - e^{- \frac{\varepsilon_{2}}{\alpha\gamma}}),
\label{eq:18}
\end{equation}
where \(\varepsilon_{1} = 216.7\ \mathrm{keV},\varepsilon_{2} = 9030\ \mathrm{keV}\) correspond to deuteron energies for which the cross section calculated from the relative deuteron--triton velocity is $0.1\,\mathrm b$. The triton energy is estimated from the simulation result that it is two-thirds of the deuteron energy.
\\

\bibliography{references}

\noindent\sfbold{Data availability}
The simulation data that support the findings of this work are available from the corresponding author upon reasonable request.
\\

\noindent\sfbold{Acknowledgements} \\
This work was supported by the Fundamental and Interdisciplinary Disciplines
Breakthrough Plan of the Ministry of Education of China (Grant No.
JYB2025XDXIM204), the Strategic Priority Research Program of the Chinese Academy of Sciences (Grant No. XDB1550100), the National Key R\&D Program of China (2025YFF0515103), and the National Natural Science Foundation of China (No. 11921006 and No. 12605382). The simulations are supported by the High-Performance Computing Platform of Peking University.
The code EPOCH is funded by UK EPSRC Grants No. EP/G054950/1, No. EP/G056803/1, No. EP/G055165/1 and No. EP/M022463/1.
Z.G. acknowledges the National Natural Science Foundation of China (NNSFC) Grants  No. 12675316 and No. 12447101, the CAS Project for Young Scientists in Basic Research (Grant No. YSBR-141), and the HPC Cluster of ITP-CAS for providing computational resources.
R.H. acknowledges the Sichuan Science and Technology Program (2025ZNSFSC0827) and the Key Research and Development Support Program of Chengdu Municipal Science and Technology Bureau (2026-YF08-00172-GX)
\\

\noindent\sfbold{Author contributions} \\
G.Q. and Z.G. conceived the research. G.Q. carried out the simulations. G.Q. and Z.G. conducted the data analysis and theoretical analysis. The manuscript was written by G.Q. with feedback and important contributions from all authors. Z.G. and X.Y. supervised and coordinated the project.
\\

\noindent\sfbold{Competing interests} \\
The authors declare no competing interests.
\end{document}